\documentclass[aps,prd,superscriptaddress,nofootinbib,showpacs,%
  twocolumn,preprintnumbers,amsmath,amssymb]{revtex4-1}
\usepackage{graphics}
\usepackage[dvips]{graphicx}
\usepackage[usenames]{color}
\usepackage[colorlinks=true,linktocpage=true,%
  linkcolor=blue,citecolor=blue,urlcolor=blue]{hyperref}

\newcommand \beq{\begin{eqnarray}}
\newcommand \eeq{\end{eqnarray}}
\def\simge{\mathrel{%
       \rlap{\raise 0.511ex \hbox{$>$}}{\lower 0.511ex \hbox{$\sim$}}}}
\def\simle{\mathrel{
       \rlap{\raise 0.511ex \hbox{$<$}}{\lower 0.511ex \hbox{$\sim$}}}}

\newcommand{\tr}{\mathop{\mathrm{tr}}}
\newcommand{\DCFL}{\Delta_{\text{CFL}}}
\newcommand{\nuA}{\nu_{_{\text{A}}}}
\newcommand{\nuB}{\nu_{_{\text{B}}}}

\newcommand{\nuON}{\nu_1}
\newcommand{\nuTW}{\nu_2}
\newcommand{\nuTH}{\nu_3}

\newcommand{\muq}{\mu_{\text{q}}}
\newcommand{\muB}{\mu_{_{\text{B}}}}

\newcommand{\calC}{\mathcal{C}}

\newcommand{\eqn}[1]{(\ref{#1})}

\usepackage[normalem]{ulem}  

\begin{document}
\title{Continuity of vortices from the hadronic
  to the color-flavor locked phase\\ in dense matter}

\author{Mark G. Alford}
\affiliation{Department of Physics, Washington University,
  St Louis, MO 63130, USA}

\author{Gordon Baym}
\affiliation{Department of Physics, University of Illinois,
  1110 W.\ Green Street, Urbana, IL 61801-3080, USA}
\affiliation{iTHES Research Group and iTHEMS Program, RIKEN, Wako,
  Saitama 351-0198, Japan}

\author{Kenji Fukushima}
\affiliation{Department of Physics, The University of Tokyo,
  7-3-1 Hongo, Bunkyo-ku, Tokyo 113-0033, Japan}

\author{Tetsuo Hatsuda}
\affiliation{iTHES Research Group and iTHEMS Program, RIKEN, Wako,
  Saitama 351-0198, Japan}
\affiliation{Nishina Center, RIKEN, Wako, Saitama 351-0198, Japan}

\author{Motoi Tachibana}
\affiliation{Department of Physics, Saga University,
  Saga 840-8502, Japan}

\date{01 October 2018}


\begin{abstract}
We study how vortices in dense superfluid hadronic matter can connect to vortices in superfluid quark matter, as in rotating neutron stars, focusing on the extent to which quark-hadron continuity can be maintained.   As we show, a singly quantized vortex in three-flavor symmetric hadronic matter can connect smoothly to a singly quantized non-Abelian vortex in three-flavor symmetric quark matter in the color-flavor locked (CFL) phase, without the necessity for boojums appearing at the transition.
\end{abstract}

\preprint{RIKEN-QHP-367}

\maketitle

\section{Introduction}
\label{sec:intro}


    In a rotating neutron star, the superfluid components -- the nuclear liquid at lower densities and a possible
color-flavor locked (CFL) quark phase~\cite{Alford:1998mk} at higher densities in the interior -- carry angular momentum in the form of quantized vortices.
How, we ask, are the vortices in these two phases connected?  Can one have
continuity or must there be a discontinuity?  How do the possible connections depend on the 
particular flavor structure of the matter?   In the ground state of dense matter, the picture of quark-hadron continuity~\cite{Schafer:1998ef,Alford:1999pa} is that as the baryon density is increased matter undergoes a smooth crossover from the hadronic phase to the quark phase.  By studying how such vortices connect we can shed
further light on whether the notion of quark-hadron continuity can be extended to angular momentum carrying states of dense hadronic
matter. 

    To summarize the problem in matching hadronic with CFL vortices we note that superfluid vortices in the BCS-paired hadronic phase have quantized circulation, $C_{\rm B}$, i.e.,
\beq  
 C_{\rm B} = \oint_{\calC} \vec v \cdot d\vec{\ell} = 2\pi \frac{\nuB}{2\muB},
  \label{circ}
\eeq
where the contour $\calC$ of integration encircles the vortex, $\muB$ is the baryon chemical potential, and $\nuB$ is an integer.  We detail this result further below.   (We work in units $\hbar$ = $c$ =1.)   All but singly
quantized vortices ($\nuB = \pm 1$) are unstable.    In a BCS-paired CFL quark phase on the other hand,  the simple 
Abelian vortex~\cite{Forbes:2001gj,Iida:2002ev}, the analog of the hadronic vortex, has circulation~\cite{Iida:2001pg}
\beq
 C_{\rm A} = \oint_{\calC} \vec v \cdot d\vec{\ell} = 2\pi \frac{\nuA}{2\muq},
  \label{circq}
\eeq
where $\muq = \muB/3$ is the quark chemical potential, and again $\nuA$ is an integer.    Singly quantized $U(1)_{\rm B}$ Abelian vortices in the quark phase have three times the circulation of singly quantized hadronic vortices.  

\begin{figure}
  \includegraphics[width=0.8\columnwidth]{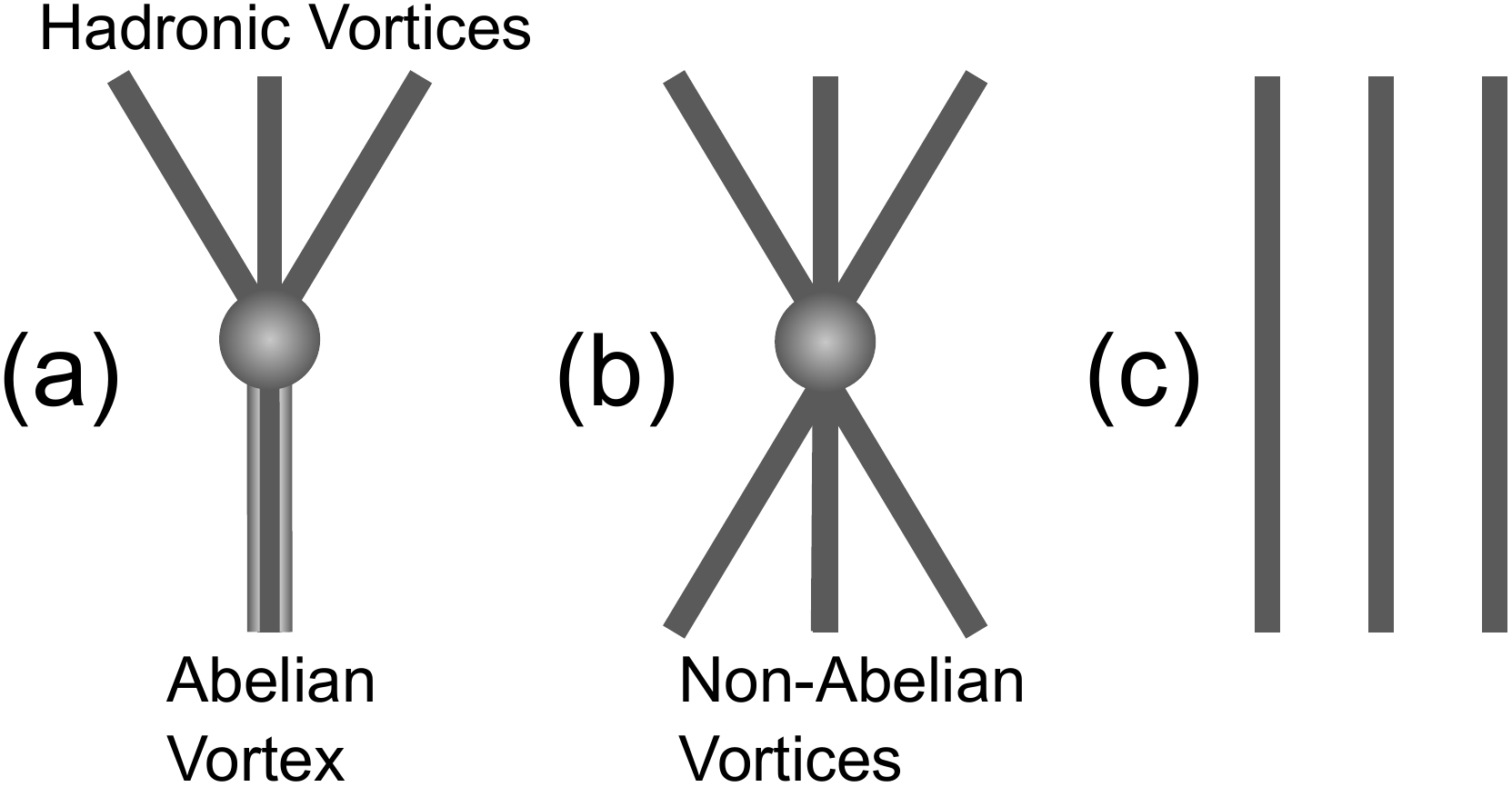}
  \caption{Schematic illustrations for connecting vortices:
    (a) If angular momentum in the CFL phase is carried by Abelian CFL
vortices then in the crossover to the hadronic phase a
``boojum'' (shaded circle) joins three hadronic vortices to a single
Abelian CFL vortex;
    (b) because Abelian CFL vortices are unstable, three hadronic vortices match onto three non-Abelian CFL vortices through
a modified boojum; or
  (c) each hadronic vortex matches onto a single non-Abelian CFL vortex without the need for a boojum.}
  \label{fig:scenarios}
\end{figure}

   Thus if one were to imagine a singly
quantized hadronic vortex turning into a singly quantized Abelian CFL vortex, the baryon velocity would have to jump discontinuously
by a factor of three from the hadronic to the quark phase, eliminating any possibility of quark-hadron continuity.  Indeed, to make the velocity continuous one would have to join three hadronic vortices to a single Abelian quark vortex, as illustrated in Fig.~\ref{fig:scenarios}(a).  Such a join is known as a ``boojum''~\cite{Mermin1977}.

    Single Abelian vortices in the CFL phase, however, are unstable against separating into three
non-Abelian vortices~\cite{Balachandran:2005ev,Nakano:2007dr,Alford:2016dco}, each of which has 1/3 the circulation of the Abelian vortex.\footnote{In Ref.~\cite{Balachandran:2005ev} these configurations were referred to as
``semi-superfluid strings,'' however we will call them ``non-Abelian vortices''
to emphasize the presence of non-Abelian color magnetic flux in the core
combined with vortex-like global rotation of the quark condensate.}
 Thus one might envisage a join with a continuous baryon velocity, as shown in 
Fig.~\ref{fig:scenarios}(b), where a boojum connects three hadronic vortices with three non-Abelian CFL vortices~\cite{Cipriani:2012hr,Eto:2013hoa}.  However, as we discuss in this paper, one does not have to make a join involving three vortices in the hadronic phase, but rather one can make a baryon-velocity conserving join between a single hadronic vortex and a single non-Abelian vortex in the CFL phase, as shown in Fig.~\ref{fig:scenarios}(c), without any need for a boojum.  To the extent that the various flavor quantum numbers permit a smooth transition from the hadronic to the CFL quark phase, angular momentum carrying states remain consistent with quark-hadron continuity.  

   To spell out this picture in detail, we first discuss more precisely the nature of quark-hadron continuity between the
hadronic and quark phases.   On the deconfined quark side the (ideal) CFL phase contains
$u$ (up), $d$ (down), and $s$ (strange) quarks, all with the same mass,  with a Fermi sea equally
populated with all three flavors and all three colors of quarks.    The corresponding hadronic phase,  three-flavor hyperonic matter,
contains all members of the light baryon flavor octet  --
$n$, $p$, $\Lambda$, $\Sigma^0$, $\Sigma^\pm$, $\Xi^0$, and $\Xi^-$ -- all of the same mass.  In the ground state at finite density, the particles populate a Fermi sea with all states of the octet equally present.

    Both phases break chiral symmetry~\cite{Alford:1998mk} and $U(1)_{\rm B}$, with the same symmetry breaking pattern [$SU(3)_{\rm L}\otimes SU(3)_{\rm R}\otimes U(1)_{\rm B} \to SU(3)_{\rm V}$].  
In the hadronic phase, the dibaryon condensate, which breaks $U(1)_{\rm B}$, is formed from two paired flavor octets, while in the
CFL phase, a diquark condensate is formed, which in the unitary gauge has the same color-flavor orientation everywhere.\footnote{With full three-flavor symmetry, CFL pairing is the most stable~\cite{Hong:2003zq,Alford:2007xm}.}
Also, in the hadronic phase, chiral symmetry is
spontaneously broken by a quark-antiquark chiral condensate, producing a light octet of pseudoscalar mesons,
i.e., $\pi^0$, $\pi^\pm$, $K^0$, $\bar{K}^0$, $K^\pm$, and $\eta$, while in the  CFL phase, the
diquark condensate spontaneously breaks chiral symmetry, producing a light octet of pseudoscalar mesons~\cite{Son:1999cm,Son:2000tu,Rho:1999xf}.
Previous studies~\cite{Schafer:1998ef,Alford:1999pa,Fukushima:2004bj,Hatsuda:2008is}
have established the continuity between the low-energy excitations of such
three-flavor hadronic and three-flavor quark matter.\footnote{This continuity is 
an example of the complementarity
between the confined and Higgs phases of a non-Abelian gauge theory \cite{Fradkin:1978dv}.}
The nine single-quark excitations of different colors and flavors
can be mapped, in the unitary gauge, onto the baryon
octet plus a baryon singlet which is usually not mentioned in
discussions of the confined phase because it is much heavier than
the octet baryons~\cite{Alford:1999pa}.

    One can further understand quark-hadron continuity in terms of
the anomaly-induced coupling between the chiral and diquark
condensates~\cite{Hatsuda:2006ps,Abuki:2010jq}.
The implications of quark-hadron continuity for
the QCD phase diagram are reviewed in Ref.~\cite{Fukushima:2010bq},
and for neutron stars in Ref.~\cite{Baym:2017whm}.

\begin{figure}
\includegraphics[width=0.5\columnwidth]{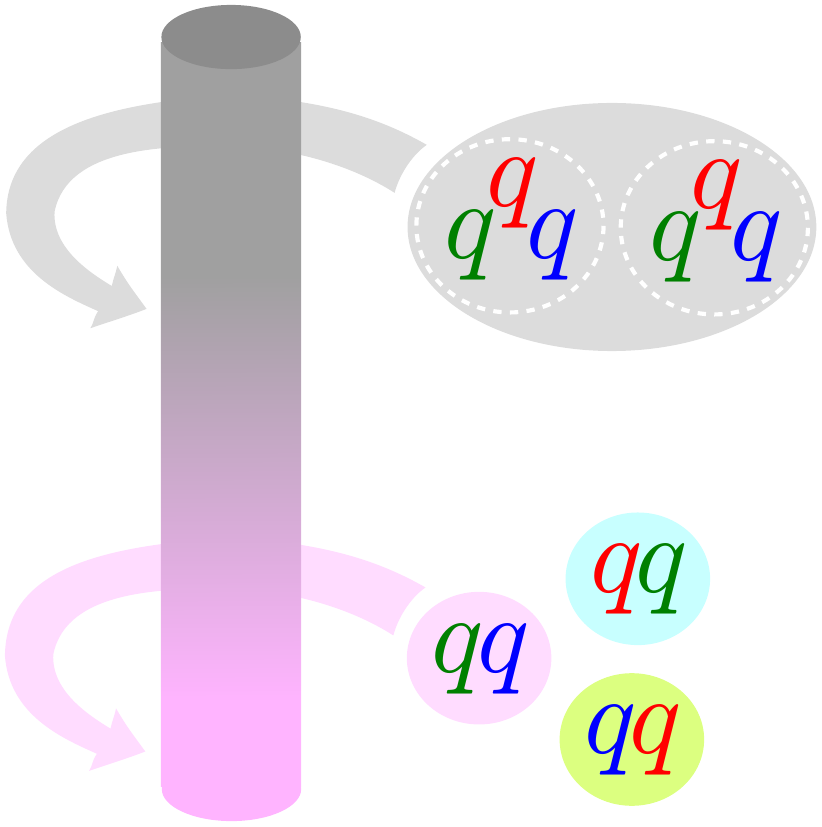}
\caption{
Schematic illustration of the smooth evolution of a
hadronic vortex into a non-Abelian CFL vortex.
In the hadronic phase, the phase of the 
condensate corresponding to paired baryons (six quarks) increases by $2\pi$ in 
winding around the vortex core.   In the CFL phase in the gauge-fixed picture, one component of the order parameter picks up a phase 2$\pi$
in winding, as shown.  In the gauge-invariant picture the phase of the entire six-quark order parameter changes by $2\pi$ in winding.
}
\label{fig:vortex}
\end{figure} 

    Figure~\ref{fig:vortex} summarizes our results.  In the confined phase
(upper half of the figure) the hadronic vortex
carries angular momentum via the circulation of a gauge-invariant
dibaryon condensate which acquires a phase of $2\pi$ when
transported around the core.  This vortex can be continuously
connected to a non-Abelian CFL vortex~\cite{Balachandran:2005ev}
in the CFL quark phase (lower half of the figure) where the vortex 
has the same baryon circulation, but it arises in the unitary gauge
from three diquark condensates, one of which acquires a
phase of $2\pi$ when transported around the core.   On the other hand, in the gauge-invariant picture, described in detail in Sec.~\ref{sec:gauge-invariant},  the phase increase is attributed to the entire six quark order parameter. 

This paper is organized as follows.
In Sec.~\ref{sec:vortex-quantization}
we review the generic properties of vortices in a
superfluid. In Sec.~\ref{sec:hadronic-and-quark}
we discuss the vortex configurations that exist
in three-flavor hadronic and quark matter.
After discussions of hadronic vortices in Sec.~\ref{sec:hadronic-vortices}, 
we describe two different
vortex configurations which have been constructed in three-flavor quark
matter,  Abelian CFL vortices in Sec.~\ref{sec:abelian-CFL}
and non-Abelian CFL vortices in Sec.~\ref{sec:nonabelian-CFL},
and then we show how a non-Abelian vortex can be continuously
connected with a hadronic vortex.  In  Sec.~\ref{sec:gauge-invariant} we show how 
these non-Abelian vortices can be understood in a gauge-invariant
description, 
focusing in Sec.~\ref{sec:flavored-vortex} on the continuity of flavored vortices.
Finally, in Sec.~\ref{Colorflux} we discuss the role of color magnetic flux.
We focus throughout on the properties of connecting single vortices, and leave the discussion of 
an array of vortices in the CFL phase at finite rotation for the future.

\section{Vortex quantization and circulation}
\label{sec:vortex-quantization}

We first review the basics of vortex quantization, circulation, and angular momentum which are common to all the vortices we
discuss here: hadronic vortices, Abelian CFL vortices, 
and CFL vortices carrying non-Abelian color flux.   

Quantized vortices arise in superfluids under rotation. 
A superfluid can be described
by a complex scalar field; the ground state expectation value 
$\Phi(\vec{r},t\/)$ of the field, in the conventional description in terms of broken symmetry, represents the condensate of bosons 
(or Cooper pairs of fermions) that gives rise to superfluidity.
The Hamiltonian for the field is invariant under
a global $U(1)$ symmetry, so that 
the number of bosons or fermions is conserved by the dynamics. 
However, if $\Phi$ is nonzero then 
the ground state of the Hamiltonian
spontaneously breaks the $U(1)$ symmetry.

In general, the condensate can be written in terms of its
modulus and phase $\phi$ as,
\beq
\Phi = e^{i\phi} |\Phi| \,.
\eeq
In the local rest frame of the condensate,
\beq
\phi= -\mu_{\rm s} t \,,
\label{mu0}
\eeq
where $\mu_{\rm s}$ is the chemical potential of the conserved particles 
in the ground state,  namely the minimum energy required to add one  boson or one pair of fermions
to the system.  Boosting to a frame in which the
condensate is in uniform motion~\cite{Baym1968}, we find
\beq
  \phi = p_\nu x^\nu = \vec p\cdot \vec r - \mu t \,,
  \label{mup}
\eeq
where $p_\nu p^\nu = - \mu_s^2$ and
$\mu=\gamma(v)\mu_s$ with $\gamma(v)\equiv 1/\sqrt{1-v^2}$.
The superfluid velocity is simply
\beq
\vec v = \frac{\vec p}{|p_0|} = \frac{\vec p}{\mu}\,.
\label{v-super}
\eeq
We can thus write the momentum carried by the unit of conserved charge
and the chemical potential as
\beq
\vec p = \vec \nabla \phi(\vec r,t) \,,\qquad
\mu = -\frac{\partial \phi(\vec r,t) }{\partial t}
\label{grad_phase}
\eeq 
for general space-time dependent $\phi$.

For a static superfluid vortex, $\phi(\vec r,t) = \phi(\vec r) - \mu t$; thus
\beq
  \Phi(\vec r\,) =   e^{i\phi(\vec r\,)-i\mu t}|\Phi(\vec r\,)| \,,
\eeq
where $|\Phi(\vec r\,)|$ is zero at the center of the vortex
and in uniform density matter is independent of position 
well outside the vortex core. Far from
the vortex core
the only spatial variation is in the phase $\phi(\vec r\,)$.

For the mathematically simplest vortex aligned along the $z$ axis,  
$\phi = \nu\varphi$, where $\varphi$ is the azimuthal angle. 
Thus the momentum per particle or pair is
\beq
\vec p\,(r) = \vec\nabla \phi = \dfrac{\nu}{r}\hat \varphi \,
\label{mom-per-charge}
\eeq
where $r$ is the distance from the vortex core and
$\hat{\varphi}$ is a unit vector in the $\varphi$ direction.  
From Eq.~\eqref{v-super} the superfluid velocity is 
\beq
v(r) = \dfrac{\nu}{\mu r}\hat \varphi \,.
\label{vortex_velocity}
\eeq
Integrating $\vec p$ along a closed contour $\cal C$ surrounding the vortex we 
obtain the total change  $\Delta \phi$ in the phase,
\beq
  \Delta\phi
  = \oint_{\cal C} \vec p \cdot d\vec{ \ell} = 2\pi\, \nu \,.
  \label{quant}
\eeq
In a three dimensional system, the winding number $\nu$ must be an integer.
From Eqs.~\eqref{v-super} and \eqref{quant} [or from Eq.~\eqref{vortex_velocity}]
the superfluid velocity obeys the circulation condition,
\beq
 C= \oint_{\cal C} \vec v \cdot d\vec{ \ell} = 2\pi \frac{\nu}{\mu},
  \label{quant1}
\eeq
as mentioned in the introduction.

   Lastly we compute the angular momentum, $L_z$, of a vortex 
centered on the $z$ axis. From Eq.~\eqref{grad_phase} the local
azimuthal momentum density is $p_\varphi n$ where $n$ is the  
particle density (as distinguished from the condensate density), 
which is independent of $\varphi$.
Thus
\beq
    L_z = \int d^3 r\, r p_\varphi\, n(r) 
 = \nu \int 2\pi r dr dz \,n(r) = N\nu \,,
\label{ang-mom-generic}
\eeq
where $N$ is the total number of particles or pairs.  
The angular momentum per particle for bosons or per fermion pair
is simply $\nu$, the winding number of the vortex.

\section{Vortices in hadronic and CFL quark matter}
\label{sec:hadronic-and-quark}

We now consider the circulation and the angular momentum associated
with vortices in hadronic and CFL quark matter.

\subsection{Hadronic vortices}
\label{sec:hadronic-vortices}

In the $SU(3)$ classification,  baryon pairs can be decomposed into irreducible representations as, 
\beq 
{\bf 8} \otimes {\bf 8} 
 = \underbrace{{\bf 1} \oplus {\bf 8} \oplus {\bf 27}}_{\mbox{sym}} ~ 
  \oplus ~ \underbrace{   {\bf 8} \oplus {\bf 10} \oplus  {\bf 10}^*   }_{\mbox{anti-sym}} \;.
\label{eq:baryon8x8}
\eeq  
Here and below, ``sym'' and ``anti-sym'' stand for the symmetry under the
flavor exchange of two baryons.
The baryon-baryon interaction in the $SU(3)$ limit is most attractive in the flavor-singlet channel ($\boldsymbol{1}$ representation)~\cite{Inoue:2011ai}
with a pairing gap of the form, 
$\Delta_{\rm B}^{(\boldsymbol{1})} =  \bigl\langle
-\sqrt\frac{1}{8} [\Lambda\Lambda]_{\rm sym} + \sqrt\frac{3}{8} [\Sigma \Sigma]_{\rm sym} + \sqrt\frac{4}{8} [N \Xi]_{\rm sym}\bigr\rangle$.
In the ground state of three-flavor hyperonic matter, flavor non-singlet pairings in other attractive channels 
can coexist with the flavor-singlet pairing, e.g., the standard nucleon pairing  in the spin-singlet isospin-triplet channel,
$\Delta_{\rm B}^{(\boldsymbol{27})} = \langle [NN]_{\rm sym} \rangle$,  and the possible
pairing in the spin-singlet isospin-doublet channel,
$\Delta_{\rm B}^{(\boldsymbol{8}_{\rm sym})} =  \langle   -\frac{1}{10} [N\Lambda]_{\rm sym}
  + \sqrt\frac{9}{10} [N\Sigma]_{\rm sym} \rangle$ \cite{Inoue:2010hs}.

In any of these pairings, the chemical potential entering Eq.~\eqref{quant1} is 
$2\muB$, that of a pair of baryons.  Therefore, no matter whether it is flavor singlet or non-singlet,
a hadronic vortex with winding number $\nuB$ has circulation $2\pi \nuB/(2\muB)$, Eq.~\eqref{circ}.
The corresponding angular momentum per baryon is [see Eq.~\eqref{ang-mom-generic}]
\beq
  \dfrac{L_{{\rm B}z}}{N_{\rm B}} = \frac{1}{2} \nuB,
   \label{Lz-hadronic}
 \label{eq:ang-B}  
\eeq
since there are $N_{\rm B}/2$ pairs in the system.

\subsection{Abelian CFL Vortices}
\label{sec:abelian-CFL}

The order parameter of quark matter in the CFL phase in the unitary gauge can be 
written in terms of the color and flavor triplet diquark operator~\cite{Alford:1998mk}
\beq
  \hat\Phi^{\alpha i} = \mathcal{N} \epsilon^{\alpha\beta\gamma}
  \epsilon^{ijk}\, q_{\beta j} C\gamma_5 q_{\gamma k} \,,
  \label{eq:UoneBop}
\eeq
where $C=i\gamma_0\gamma_2$ is the charge conjugation operator, and Greek and Latin letters denote color and flavor indices,
respectively; $\cal N$ is a normalization constant.
The order parameter is then
\beq
   \Phi^{\alpha i} = \langle \hat\Phi^{\alpha i} \rangle \,.
  \label{eq:diquark-field} 
\eeq
The matrix $\Phi^{\alpha i}$ can be diagonalized by a combination
of color and flavor rotations, so that without loss of generality we write
\begin{equation}
 \Phi =
 \begin{pmatrix}
 \Phi^{\bar r \bar u} & 0 & 0 \\ 0 & \Phi^{\bar g \bar d} & 0 \\ 0 & 0 & \Phi^{\bar b \bar s} 
  \end{pmatrix} \,,
  \label{eq:CFL}
\end{equation}
where $r, g, b$ ($\bar{r}, \bar{g}, \bar{b}$) denote colors
(anti-colors) and $u, d, s$ flavors;
in the ground state, $\Phi^{\bar r \bar u} = \Phi^{\bar g \bar d} = \Phi^{\bar b \bar s} = \DCFL$.

Naively one would expect the angular momentum carrying states with lowest
energy per unit of angular momentum, to be  global $U(1)_{\rm B}$ or ``Abelian CFL'' vortices.
In these vortices
each of the three non-zero components of the order parameter winds 
around the core of the vortex, so for an Abelian 
CFL vortex aligned along the $z$ axis 
the order parameter assumes the form
\begin{equation}
 \Phi^{\rm A} = \DCFL\,e^{i \nuA\varphi}
 \begin{pmatrix}
    f(r) & 0 & 0 \\ 0 & f(r) & 0 \\ 0 & 0 & f(r)
  \end{pmatrix} \,,
\label{eq:abelian}
\end{equation}
where $f(r)$ varies monotonically from zero at $r=0$ to unity as
$r\to\infty$, with $\nuA$ the winding number 
of the Abelian CFL vortex. 

The quark chemical potential is $\mu_{\rm q}=\muB/3$, and thus the chemical
potential per quark pair is $2\mu_{\rm q}\ = \frac23 \muB$, so from
Eqs.~\eqref{v-super} and
\eqref{grad_phase} and the total momentum per quark pair in the condensate is
\beq
  \vec{p} = \frac{2}{3}\muB\,\vec{v}\,,
  \label{eq:PA}
\eeq
where as before $\vec{v}$ is the superfluid velocity, so the circulation is
\begin{equation}
  C^{\rm A} =   \frac{3\nuA}{2\muB}
  \oint d\vec{\ell}\cdot \vec{\nabla}\varphi
  = \frac{3 \nuA}{2} \cdot \frac{2\pi}{\muB}\,.
  \label{circ-Abelian}
\end{equation}
The angular momentum per baryon of the vortex is 
\beq
  \dfrac{L_{{\rm A}z}}{N_{\rm B}} = \frac32 \nuA \,.
  \label{Lz-Abelian}
\eeq

We now ask how the vortices in hadronic matter would match on to
Abelian vortices in CFL quark matter at a crossover between these phases.
If the superfluid velocity, and hence
the circulation, Eq.~\eqref{quant1},
and angular momentum per baryon, Eq.~\eqref{ang-mom-generic},
do not match in the two phases, then quark-hadron continuity would be violated.   By comparing
Eqs.~\eqref{circ} and \eqref{circ-Abelian},
or equivalently \eqref{Lz-hadronic} and \eqref{Lz-Abelian},
we see that matching would require
\begin{equation}
  \nuB = 3\nuA \,.
  \label{eq:quant}
\end{equation}
The matching relation \eqn{eq:quant} implies that
three singly quantized
hadronic vortices should merge into one Abelian CFL vortex, violating quark-hadron continuity in states with finite angular momentum. 
This merging would require a boojum \cite{Mermin1977}
at the interface between the two phases,
as sketched in Fig.~\ref{fig:scenarios}(a).
As we discuss in the next section, the violation need not be present for the more stable non-Abelian vortices in the CFL phase.

\subsection{Non-Abelian CFL Vortices}
\label{sec:nonabelian-CFL}

   An Abelian CFL vortex
is energetically unstable against formation of 
three ``non-Abelian'' vortices~\cite{Balachandran:2005ev,Nakano:2007dr}.
The condensate of the anti-red--anti-up ($\bar{r} \bar{u}$)
non-Abelian vortex is
\begin{align}
 \Phi^{(1)} &= \DCFL \begin{pmatrix}
    e^{i \nuON \varphi} f(r) & 0 & 0 \\
    0 & g(r) & 0 \\
    0 & 0 & g(r)
 \end{pmatrix} \,,
 \label{eq:R}
\end{align}  
with corresponding gluon field
\begin{align}
  A^{(1)}_{\varphi} &= - \frac{\nuON}{g_{\rm c} r} \left[1-h(r)\right] 
 \begin{pmatrix}-\frac{2}{3} & 0 & 0 \\
  0 & \frac{1}{3} & 0 \\
  0 & 0 & \frac{1}{3} \end{pmatrix} \, \label{eq:non-Abelian-A} ,
\end{align}
where $g_{\rm c}$ is the QCD coupling and the 
boundary conditions are  
\begin{equation}
\begin{array}{rcl}
f \to 0,\quad  g'\to 0,\quad  h\to 1  & \text{~as~} & r \to 0 \,, \\
f \to 1,\quad  g \to 1,\quad  h\to 0  & \text{~as~} & r \to \infty \,.
\end{array}
\label{eq:boundary}
\end{equation}    
Single-valuedness of the condensate requires that $\nuON$ be an integer.
Anti-green--anti-down ($\bar g \bar d$) and
anti-blue--anti-strange ($\bar b \bar s$) versions,
$ \Phi^{(2)} $ with $\nuTW$ and $ \Phi^{(3)} $ with $\nuTH$,
can be obtained by permuting the diagonal elements.

  To obtain the superfluid velocity
and angular momentum per baryon of the non-Abelian vortex,
we rewrite Eq.~\eqref{eq:R} as
\begin{equation}
  \Phi^{(1)} = \DCFL e^{\frac{i}{3}\nuON\varphi}
  \left(\begin{array}{c@{\!\!}c@{\!\!}c}
    e^{\frac{2i}{3}\nuON\varphi} f(r) & 0 & 0 \\
    0 & e^{-\frac{i}{3}\nuON\varphi} g(r) & 0 \\
    0 & 0 & e^{-\frac{i}{3}\nuON\varphi} g(r)
  \end{array}\right) \,.
\label{eq:R1}
\end{equation}
In this form the overall factor of $e^{\frac{i}{3}\nuON\varphi}$ 
is the $U(1)_{\rm B}$ phase, while the phase factors within the matrix 
are a color rotation.  [We note for later computation of the covariant derivative of $  \Phi^{(1)}$ that the gradients of these phases are compensated by the color gauge field
\eqref{eq:non-Abelian-A}.]

  The chemical potential per quark pair is $2\muq = \frac23 \muB$, so from
Eqs.~\eqref{v-super}, \eqref{grad_phase}, and \eqref{mom-per-charge}
the total momentum per quark pair is related to the superfluid
velocity $\vec v$ by
\begin{equation}
  \vec{p} = \frac{1}{3}\cdot \frac{\nuON}{r}\hat\varphi
  = \frac{2}{3} \muB \vec{v}\,.
\end{equation}
The circulation around the vortex, Eq.~\eqref{quant1}, is
\begin{equation}
  C_{(1)} = \oint_{\cal C} \vec v \cdot d\vec{ \ell}
 = \frac{\pi \nuON}{\muB}\,.
  \label{eq:circularR}
\end{equation}
Correspondingly,  the angular momentum per baryon of the vortex of the form~\eqref{eq:R} or \eqref{eq:R1} is
 \begin{equation}
  \frac{L_{(1)z}}{N_{\rm B}} = \frac{1}{2} \nuON \,.
 \label{eq:ang-U} 
\end{equation}
The same relations also hold for $\Phi^{(2)}$ with $\nuTW$ and $\Phi^{(3)}$ with $\nuTH$.
 
We see from Eqs.~\eqref{circ} and \eqref{eq:circularR} and from Eqs.~\eqref{eq:ang-B} and \eqref{eq:ang-U} 
that singly quantized ($\nuB=1$) vortices in hadronic matter can match onto
singly quantized (${\nuON}=1$, ${\nuTW}=1$, or ${\nuTH}=1$) non-Abelian vortices in CFL quark matter at a crossover
between these phases, with no discontinuity in baryon velocity and angular momentum.

This result can be understood intuitively as follows.
In the hadronic vortex, the dibaryon condensate
acquires a phase of $2\pi$ as one follows it along a contour encircling the
vortex core. Since the dibaryon can be viewed as 3 diquarks, this corresponds
to each diquark acquiring a phase of $2\pi/3$.
The non-Abelian vortex in the CFL condensate has exactly the same
circulation: each diquark acquires a phase\footnote{If $U(1)_{\rm B}$
were a local gauge symmetry, the vortex would become a $U(1)_{\rm B}$ flux tube.
The hadronic vortex and the non-Abelian vortex would both have the
same $U(1)_{\rm B}$ flux in their cores.} of $2\pi/3$.

We conclude, in agreement with Ref.~\cite{Cipriani:2012hr}, that
a single non-Abelian CFL vortex has the same circulation as
a hadronic vortex.  However, Ref.~\cite{Cipriani:2012hr}
suggests that, in order to neutralize the color flux contained in the non-Abelian vortices, three non-Abelian CFL vortices must merge to form
a boojum at the CFL-hadronic boundary
to which three hadronic vortices then connect [see Fig.~\ref{fig:scenarios}(b)].
As we argue below, there is no need for such 
a boojum: a single non-Abelian CFL vortex 
can smoothly evolve into a single hadronic vortex [as in Fig.~\ref{fig:scenarios}(c)]. 
To show this, further consideration of the flavor structure of the vortices is necessary in the hadronic and the CFL phases, as we discuss in Sec.~\ref{sec:gauge-invariant}.

\subsection{Gauge-invariant description}
\label{sec:gauge-invariant}

In Sec.~\ref{sec:hadronic-and-quark} we described the CFL condensate in the unitary gauge.
Although such a gauge-fixed description is convenient for writing down the non-Abelian vortex solution explicitly and  
showing the continuity of  the circulation and angular momentum between the hadronic phase and the CFL phase,
 it is not clear how the flavor structures in the two phases are connected.
To resolve this problem, in this section we describe 
vortices in the CFL phase 
in a gauge-invariant manner~\cite{Rajagopal:2000wf}
using diquarks
in Eqs.~\eqref{eq:UoneBop} and \eqref{eq:diquark-field}
as building blocks.
We can write down meson-like and baryon-like gauge-invariant combinations of
diquark operators,
\begin{align}
  \hat{\mathcal{M}}_i^j(\vec r\,) & \equiv \hat\Phi^{\dag}_{i\alpha}\hat\Phi^{\alpha j} \,, 
  \label{eq:Mij}\\
  \hat{\Upsilon}^{ijk}(\vec r\,) & \equiv \frac{1}{6}
   \epsilon_{\alpha\beta\gamma}\hat\Phi^{\alpha i}
  \hat\Phi^{\beta j}\hat\Phi^{\gamma k} \,.
\end{align}
We will focus on $\hat{\Upsilon}^{ijk}(\vec r\,)$ for the moment and will consider $\hat{\mathcal{M}}_i^j(\vec r\,)$
later in Sec.~\ref{sec:core-flavor}.
According to quark-hadron continuity, $\langle\hat\Upsilon^{ijk}(\vec r\,)\rangle$ is nonzero in both the CFL and hadronic phases because both phases break
baryon number, via diquark and dibaryon condensates respectively. In
Secs.~\ref{sec:singlet-vortex} and \ref{sec:flavored-vortex} below
we will discuss the projection of $\hat{\Upsilon}^{ijk}(\vec r\,)$ onto
specific flavor representations.

In the CFL phase, in the mean field approximation,
\beq
  \Upsilon^{ijk}(\vec r\,) \equiv \langle \hat\Upsilon^{ijk}(\vec r\,) \rangle
  = \frac{1}{6} 
  \epsilon_{\alpha\beta\gamma} 
  \Phi^{\alpha i}\Phi^{\beta j}\Phi^{\gamma k} \,.
\label{eq:CFL-color-s}
\eeq
$\Upsilon^{ijk}(\vec r\,)$  provides a gauge-invariant description of
the non-Abelian vortex originally defined through the gauge-dependent
condensate $\Phi$.
 
Note that the irreducible flavor $SU(3)$ decomposition of $\Upsilon^{ijk}(\vec r\,)$ is
\beq 
\boldsymbol{3}^*\otimes\boldsymbol{3}^* \otimes \boldsymbol{3}^* 
=\boldsymbol{1}\oplus
\boldsymbol{8} \oplus\boldsymbol{8} \oplus\boldsymbol{10}^* \,,
\eeq
so that not only flavor-singlet but also flavored vortices can be obtained 
from  $\Phi$ by appropriate projections.
These would match to certain of the hadronic vortices classified in Eq.~\eqref{eq:baryon8x8}.

   According to \eqn{eq:CFL-color-s}
the total number of 6-quark condensates in the CFL
phase is $3\times 3\times 3=27$, while the number of 
pairs of octet baryons in
the hadronic phase is $8\times 8=64$.  One might think that there is a
mismatch, but this is because our diquark
condensate $\Phi$ only includes flavor antisymmetric diquarks.
We will discuss this point in Sec.~\ref{sec:flavored-vortex}.

   In the hadronic phase a nonzero expectation value of $\hat\Upsilon^{ijk}(\vec r\,)$
is an order parameter for baryon number violation,
which is manifest with $\hat\Upsilon^{ijk}(\vec r\,)$ rewritten in
terms of the baryon-interpolating operator,
$\hat{B}^{i\,a}_{j}\equiv\hat\Psi^{\alpha i}\hat q_{\alpha j}^a$; the spin-$1/2$
is represented by the index $a$ on $q_{\alpha j}^a$.   In writing
$\hat{B}^{i\,a}_{j}$ as interpolating operators for spin-$1/2$ baryons,
we simplify the operator structure by neglecting the axial vector
diquark (called the ``bad diquark'' in hadron structures), which is a
reasonable approximation for low-lying baryons.  
The operator $\hat{B}^{i\,a}_{m}$ can be written as a sum of 
flavor-singlet and flavor-octet operators as
\beq
 \hat{B}^{i\,a}_{m}=\hat{B}^a_{\bf 1}(\delta^i_{m}/\sqrt{6})
+\hat{B}^{A\,a}_{\bf 8} (t^A)^i_{m},
\label{BBB}
\eeq
where the $t^A$ are the
SU(3) generators ($A=1,\dots,8$) in flavor space, with the normalization $\tr (t^A)^2 = 1/2$.
Then $\hat{B}^a_{\bf 1}\equiv 2\tr(\hat{B}^a)/\sqrt{6}$ and
$\hat{B}^{A\,a}_{\bf 8}\equiv 2\tr(t^A \hat{B}^a)$.

  Forming $\hat{B}^{i\,a}_j$ by combining the quark operator with the diquark operator written in terms  of two quarks, (\ref{eq:UoneBop}),  we find the operator relation
    \beq
    \hat\Upsilon^{ijk}(\vec r\,) = \frac{1}{3} \epsilon^{kmn}
    (C\gamma_5)_{ab}\, \hat{B}^{i\,a}_{m} \hat{B}^{j\,b}_{n} \,.
    \label{eq:Upsilon-dibaryon}
    \eeq
Clearly, a dibaryon condensate  $\langle \hat{B}\hat{B}\rangle \neq 0$ in the hadronic phase, makes
$\Upsilon^{ijk}$ nonzero.

\subsubsection{Flavor-singlet vortex}
\label{sec:singlet-vortex}

   We first consider vortices in the flavor-singlet projection of the gauge-invariant order parameter,
\beq
  \hat\Upsilon_{\bf 1}(\vec r\,) = \epsilon_{ijk}  \hat\Upsilon^{ijk}(\vec r\,)\,.
 \label{eq:CFL-ginv-cond}
\eeq
We can equivalently express this expectation value using 
  Eq.~\eqref{eq:Upsilon-dibaryon} in terms of the baryon operators, (\ref{BBB}),
\begin{align}
    \Upsilon_{\bf 1}(\vec r\,)
    &= \frac{1}{3}(C\gamma_5)_{ab}\bigl( \delta_i^{m}\delta_j^{n}
      - \delta_i^{n}\delta_j^{m} \bigr)
    \langle \hat{B}^{i\,a}_{m} \hat{B}^{j\,b}_{n}\rangle\notag\\
    &= \frac{1}{3}(C\gamma_5)_{ab}\biggl(
      \langle \hat{B}^a_{{\bf 1}} \hat{B}^b_{{\bf 1}}\rangle
      -\frac{1}{2}\langle \hat{B}^{A\,a}_{\bf 8}
      \hat{B}^{A\,b}_{\bf 8}\rangle \biggr)\,;
\label{eq:singlet-dibaryon}
\end{align}
in hadronic language
$\Upsilon_{\bf 1}(\vec r\,)$ corresponds to a flavor-singlet
condensate made with flavor-singlet and flavor-octet baryons.

In the CFL phase insertion of any of $\Phi^{(1)}$, $\Phi^{(2)}$ 
or $\Phi^{(3)}$ gives the same form
 \beq
\Upsilon_{\bf 1} = e^{i\nu_q \varphi} \DCFL^3 f(r) g^2(r),
\label{eq:UpsilonR}
\eeq  
which implies that the non-Abelian vortices $\Phi^{(1,2,3)}$ have a common flavor-singlet component.
A singly quantized ($\nu_q=1$) vortex has the same circulation $2\pi/2\mu_B$ as a singly quantized
($\nuB=1$) hadronic vortex in the flavor-singlet channel;
its phase winds by $2\pi$ on a contour encircling the vortex core, consistent with our finding that these two vortices match smoothly onto each other, with quantized vortex circulation $2\pi/2\mu_B$.

If, on the other hand, were we to substitute the field configuration for an Abelian vortex $\Phi^{({\rm A})}$ in Eq.~\eqref{eq:abelian}
into Eq.~\eqref{eq:CFL-ginv-cond}, we would find
\beq
 \Upsilon_{\rm A} = e^{3i\nuA \varphi} \DCFL^3 f^3(r) \,;
\eeq
the gauge-invariant form of a singly quantized Abelian vortex winds
three times more (by $6\pi$) on a contour encircling the vortex core.
This winding is consistent with needing three hadronic vortices to
match to one Abelian vortex~\cite{Cipriani:2012hr}.

We now consider the vortex energy in terms of the gauge-invariant order parameter.
Because of the boundary condition~\eqref{eq:boundary}, the extra energy density
of a vortex far away from its core arises from the derivative
terms;  for a non-Abelian vortex the energy density is asymptotically 
\beq
 \epsilon^{(1)}\, = \tr |\boldsymbol{D}\Phi^{(1)}|^2 \,,
\label{eq:energyR}
\eeq
where the covariant derivative is $\boldsymbol{D}=\boldsymbol\nabla -ig_{\rm c} \boldsymbol{A}$,  and the trace is taken with respect to color-flavor matrix indices.
The gluon field~\eqref{eq:non-Abelian-A} in $\boldsymbol{D}$ exactly
cancels the derivatives of the phases in the color-flavor matrix
part of $\Phi^{(1)}_{\alpha i}$ in Eq.~\eqref{eq:R1}.  As a result only
the derivative of the $U(1)_{\rm B}$ phase contributes to the energy density at large distance from the vortex core,
\beq
 \epsilon^{(1)} \,= 3\cdot \frac{\nuON\,^2}{9r^2}|\DCFL|^2 \,.
\eeq
Calculating $\nabla \Upsilon_{\bf 1}$ from Eq.~(\ref{eq:UpsilonR}) we can write the
energy in terms of the gauge-invariant order parameter as
\beq
 \epsilon_{\bf 1} \,=\frac{1}{3(\DCFL)^4}|\boldsymbol{\nabla} \Upsilon_{\bf 1}|^2\,.
\eeq
This is the kinetic term of a Ginzburg-Landau theory \cite{Iida:2000ha} at large distance for the gauge-invariant flavor-singlet order
parameter $\Upsilon_{\bf 1}$.

   We can write the full gauge-invariant Ginzburg-Landau free energy in two-dimensions in the form:
\beq
 F = \mathcal{N}\int d^2 r\, \biggl( |\boldsymbol{\nabla}\tilde{\Upsilon}_{\bf 1}|^2
  - m^2 |\tilde{\Upsilon}_{\bf 1}|^2 + \frac{\lambda}{2} |\tilde{\Upsilon}_{\bf 1}|^4 \biggr) \,,
\label{eq:singlet-LG}
\eeq
where we rescale $\Upsilon_{\bf 1} \to \tilde{\Upsilon}_{\bf 1}$
to make the coefficient of the gradient term be unity at the
mean-field level.  The full determination of the coefficients, $m^2$ and $\lambda$, from QCD is a challenging future problem.  This form of the  Ginzburg-Landau free energy  describes the interaction between the flavor-singlet parts of non-Abelian vortices (see also Ref.~\cite{Auzzi:2007wj}).

    As in simple superfluids, e.g., $^4$He, the interaction energy of two non-Abelian vortices in the gauge-invariant picture is essentially 
the integral of the product of the two vortex velocities, $\boldsymbol{v}_1\cdot \boldsymbol{v}_2$, which is generally negative between two similarly quantized vortices;  for two singly quantized vortices whose cores are separated by $L$, assumed much greater than the coherence length $1/m$, the interaction energy is\footnote{The interaction free energy of two vortices, one at the origin with phase $\phi_1$ and the second with phase $\phi_2$, where the $\phi$'s are the azimuthal angles $\varphi$ measured from the individual vortex cores, is
$F_{\rm int} = \int d^2r \boldsymbol{\nabla}\varphi_1\cdot \boldsymbol{\nabla}\varphi_2 |\tilde{\Upsilon}_0|^2$.  After integration by parts only the surface term
remains, since $\nabla^2 \varphi = 0$, and choosing the branch cut in the phase along the $x$ axis, the integral becomes
$\int_{1/m}^L dx\, \partial_y\varphi_1\cdot \Delta\varphi_2 |\tilde{\Upsilon}_0|^2.$ 
Since $\Delta\varphi_2$ (except at its core, where the order parameter vanishes), the discontinuity of $\varphi_2$ along across the $x$ axis is $ -2\pi$, we find Eq.~(\ref{vint}).}
\beq
 F_{\rm int} = -\frac{2\pi m^2}{\lambda}  \ln (m\, L)\,.
\label{vint}
\eeq
Here, the coefficient appears from the normalized condensate,
  $|\tilde{\Upsilon}_{\bf 1}|^2 = m^2/\lambda$
  in the mean-field approximation.
This logarithmically diverging result (see \cite{khalat})
indicates that the two vortices repel.

\subsubsection{Flavored vortices}
\label{sec:flavored-vortex}

   We now consider vortices in the flavor-octet projection of the gauge-invariant
order parameter,
\beq
  \hat{\Upsilon}_{\bf 8}^A =
  \epsilon_{ij\ell}(t^A)^\ell_k \hat{\Upsilon}^{ijk}(\vec r\,) \,. 
  \label{eq:CFL-g}
\eeq
This term vanishes in the mean field approximation, but beyond mean field
the flavor-octet part of non-Abelian CFL vortices could
smoothly connect to flavor-octet hadronic vortices, just as
the flavor-singlet part of a non-Abelian vortex can smoothly connect to a
flavor-singlet hadronic vortex.  
As with the flavor singlet, we can express
$\Upsilon_{\bf 8}^a (\vec r\,)$ in terms of the baryon operators,
\begin{align}  \label{eq:octet-dibaryon}
    \Upsilon_{\bf 8}^A(\vec r\,)
    &= \frac{1}{3}(C\gamma_5)_{ab}
    \epsilon_{ij\ell}(t^A)_k^\ell \epsilon^{kmn}
    \langle \hat{B}^{i\,a}_{m} \hat{B}^{j\,b}_{n}\rangle \\
    &= \frac{1}{6} (C\gamma_5)_{ab} \biggl(
      d^{ABC} \langle \hat{B}^{B a}_{\bf 8} \hat{B}^{C b}_{\bf 8} \rangle
      - \frac{\sqrt{6}}{3} \langle\hat{B}_{\bf 1}^a \hat{B}_{\bf 8}^b
      \rangle \biggr)\,,  \notag
 \end{align}
where the $d$ tensor is defined by
$\{\lambda^A,\lambda^B\} = \frac{1}{3}\delta^{AB} + d^{ABC} \lambda^C$.
Equation~\eqref{eq:octet-dibaryon} shows how the flavor octet vortex
$\Upsilon_{\bf 8}^A$ can be understood as a symmetric 
${\bf 8}$ made with two octet baryons [as classified in Eq.~\eqref{eq:baryon8x8}].

We note that the flavor structure of dibaryon pairings such as  $\langle nn\rangle$ and $\langle pp\rangle$ in
two-flavor superfluid nuclear matter cannot be realized in the present setup for the CFL phase.
 For example, a neutron pair condensate, $\langle nn\rangle$, has an overlap
with the diquark condensate, $\langle ud\rangle\langle ud\rangle\langle dd\rangle$; however, because $\langle dd\rangle$ is
flavor symmetric, it must be color symmetric for a spin-singlet
(antisymmetric) pair, and thus cannot be constructed out of
$\Upsilon^{ijk}$ given in terms of $\Phi^{\alpha i}$.
Such pairing is possible in the color sextet channel; although single gluon exchange is
repulsive for
color-triplet diquarks, and such pairing is presumably less favored, nonetheless this pairing breaks the same symmetries 
and is therefore induced by color antisymmetric pairing
\cite{Alford:1999pa,Pisarski:1999cn}.   Another possible way to form $\langle dd\rangle$ is with color-triplet
and spin-triplet pairing~\cite{Alford:2002rz,Schmitt:2004et}, which has spin one and breaks rotational symmetry.
Such states could connect naturally to $^3 P_2$ pairing in
dense nuclear matter.   We leave the question of vortex continuity between neutron  $^3 P_2$ pairing and color-triplet.
spin-triplet paired quark matter for the future.

\subsubsection{Flavor symmetry breaking in the vortex core}
\label{sec:core-flavor}

At least at the level of the mean-field approximation, flavor symmetry
is spontaneously broken in the core of a CFL vortex~\cite{Eto:2013hoa},
$SU(3) \to SU(2)\otimes U(1)$, which can be characterized by 
the flavor-octet order parameter
$\mathcal{M}_i^j=\langle\mathcal{\hat M}_i^j\rangle$ introduced in
Eq.~\eqref{eq:Mij}.  For a $\Phi^{(3)}$ condensate, for example,
we have 
\beq
  \tr(t^{A} \mathcal{M}) = -\dfrac{2}{\sqrt{3}}
    \bigl[ f(r)^2 \!-\! g(r)^2 \bigr]\, \delta^{A,8}.
\eeq
Whether this prediction survives beyond mean field requires analysis
of the fluctuation modes of a CFL vortex in (3+1) dimensions.  If the
core is effectively a (1+1) dimensional system, the
Mermin-Wagner-Hohenberg-Coleman
theorem~\cite{Mermin:1966fe,Hohenberg:1967zz,Coleman:1973ci} would
imply that fluctuations in the order parameter along the symmetry
broken directions (the CP$^2$ mode~\cite{Eto:2013hoa}) would prevent
spontaneous breaking of continuous symmetries in systems in
(1+1) dimensions at $T\geqslant 0$ [and in (2+1) dimensions at
  $T > 0$ \cite{Ma:1974tp}].
This indicates that if the Hamiltonian is flavor symmetric no
flavor-breaking condensate would be able to appear in the vortex
core.  (See Ref.~\cite{Nitta:2013wca} and references therein for
detailed discussions on the absence of flavor symmetry breaking in the
vortex core in relativistic theories as well as possible exceptions in
non-relativistic theories.)
Even if flavor symmetry is spontaneously broken in the CFL vortex,
e.g., due to a coupling between the Kelvin mode and the CP$^2$ mode
which requires (3+1)-dimensional analysis, the octet components of
$\hat\Upsilon^{ijk}$ could develop an expectation value inside the
hadronic vortex core.

Therefore, in either scenario, the flavor transformation properties of
the CFL vortices do not prevent continuity of vortices between the
hadronic and CFL phases.

\section{Color flux}
\label{Colorflux}

In Sec.~\ref{sec:hadronic-and-quark} we argued that at a crossover
between the hadronic phase and the CFL phase, a hadronic vortex
can smoothly evolve into a non-Abelian CFL vortex, in keeping with
quark-hadron continuity.
More generally, even if there is a first order phase transition
between the CFL and hadronic phase (terminating a CFL vortex in much
the same way as vortex terminates at a free surface in a liquid), it
is hard to avoid a hadronic vortex, since then one would have to have
a layer of discontinuity in the baryonic current.  This raises the
question of what happens to the color magnetic flux in the non-Abelian
CFL vortex.   Reference~\cite{Cipriani:2012hr} argued that
at the quark-hadronic boundary there must be a boojum
where three non-Abelian CFL vortices 
with different color magnetic fluxes
come together so that their color fluxes cancel,
and they can then connect to three hadronic vortices
[see Fig.~\ref{fig:scenarios}(b)].
However, we argue that there is no need for such an elaborate
construction. 

The gauge-invariant characterization of the color magnetic flux was
recently discussed in Ref.~\cite{Cherman:2018jir} which noted that, 
just as for local
non-Abelian flux tubes \cite{Preskill:1990bm,Alford:1990fc}, 
the color magnetic flux
in the non-Abelian CFL vortex
can be characterized by the Aharonov-Bohm phase
acquired by a heavy ``probe'' quark when 
transported around the vortex. This is manifest in the expectation value of 
the trace of the Wilson loop operator, 
\beq
   W_3({\cal C}) = \frac1{N_{\rm c}}\tr {\cal P} \exp\!\left({ig_{\rm c}\oint_{\cal C} ds_\mu A^{\mu A}t_3^A}\right),
\eeq
where $N_{\rm c}$ is the number of colors (i.e., $N_{\rm c}=3$), $\cal P$ denotes path ordering, 
the $t_3^A$ are the $SU(3)$ color generators
in the triplet representation, and  ${\cal C}$ is a 
closed contour encircling the vortex. If the contour is large enough then
the Wilson loop follows a perimeter law
$\langle W_3({\cal C}) \rangle = \chi_{_{\cal C}} \exp(-\kappa L({\cal C}))$ 
in both phases, where $L({\cal C})$ is the length of the contour, and $\kappa$ is  
an effective mass.   The prefactor $\chi_{_{\cal C}}$  contains the Aharonov-Bohm phase
for the path ${\cal C}$, normalized so that for a large contour ${\cal C}_0$
that does not encircle a vortex, $\chi_{_{{\cal C}_0}} =1$.

Reference~\cite{Cherman:2018jir} emphasized that, for a non-Abelian CFL vortex,
$\chi_{_{\cal C}}$ is a $Z_3$ phase, an element of the center of the color
gauge group,  
whereas in the hadronic
phase we expect that for a contour ${\cal C}$ encircling a hadronic vortex there will
be no such phase, $\chi_{_{\cal C}} =1$, since there is no color flux in
the hadronic vortex.
However, as we now explain,
this does not mean that a boojum is required at the quark-hadron boundary.

One of the leading scenarios for explaining confinement is the
condensation of ``center vortices'' \cite{tHooft:1977nqb,Vinciarelli:1978kp,Yoneya:1978dt,Cornwall:1979hz,Nielsen:1979xu};
for a recent review see \cite{Greensite:2016pfc}.
According to this picture, the confining QCD vacuum is filled with flux tubes
that carry $Z_3$ color flux. It is therefore quite possible that when
a non-Abelian CFL vortex arrives at the CFL-hadronic boundary, its
color flux can leak away into the confined hadronic phase, indistinguishable from
the pre-existing condensate of center vortices. 
There is no reason why multiple CFL vortices should be constrained to
converge at a boojum before entering the hadronic phase: even if they
are far apart their
color fluxes can still cancel by connecting with each other through
the putative condensate of color vortices in the hadronic phase.

   The behavior of the color flux in the center vortex picture is well
   illustrated in a spherical compact stellar object made of
$SU(3)$-symmetric matter, rotating so slowly around the central $z$ axis that 
it contains exactly one vortex lying along this axis.   We assume that the lower-density mantle is in the
hadronic phase and the higher-density core is in the CFL phase.  The vortex has a ``southern'' hadronic 
segment, a central CFL segment,
and a ``northern'' hadronic segment,  Since such a system
cannot contain a boojum, which requires three vortices, what then is configuration of 
the color flux?      When the CFL vortex reaches the hadronic phase, at the north pole of the core, its
$U(1)_{\rm B}$ circulation becomes the northern segment of the hadronic vortex which
continues upwards along the $z$ axis, and in the center vortex picture,  its color flux would become another
member of the existing condensate of center vortices in the hadronic phase.
That color flux is redistributed through a chain of monopoles and
antimonopoles connected by color flux tubes 
\cite{Greensite:2016pfc} in
the hadronic mantle and
ultimately
links to the south pole of the core, where it would combine with the
southern segment of the hadronic vortex to create the CFL vortex that begins
at the south pole of the core.

\section{Conclusions}

   We have argued here that singly quantized superfluid vortices in three-flavor symmetric hadronic matter can transform smoothly into singly quantized non-Abelian superfluid vortices in three-flavor symmetric color-flavor locked quark matter, without the need to include boojums to mark the transition at the interface between the two phases.     
   One can make a one-to-one correspondence between vortices in the baryonic and quark phases.  We have constructed a gauge invariant description of non-Abelian vortices.  A natural next step will be to spell out the full Ginzburg-Landau theory for non-Abelian vortices in terms of their gauge-invariant order parameter.
   
    We have only studied the question of the connections of single vortices in fully $SU(3)$ flavor symmetric matter.  To make our analysis applicable to more realistic situations in neutron stars where one does not have even isospin symmetry requires extending the analysis to flavor-symmetry broken states, resulting from the higher mass of the strange quark
(for a discussion of the ramifications for CFL superfluid vortices see
Ref.~\cite{Eto:2013hoa}).  The extension will require considering BCS pairing states in the quark phase beyond ideal CFL with simple color, flavor, and spin asymmetry.  
Ultimately we would like to determine the extent to which one can
connect the hadronic and quark matter phases and their vortices in a smooth way. 
Furthermore, at large rotational rates one expects an array of vortices.  While in the hadronic phase the vortices are expected to form a triangular lattice, to determine the optimal lattice configurations in the quark phase requires better understanding the interactions of non-Abelian vortices.

\section*{Acknowledgments}
The authors thank Aleksey~Cherman, Muneto~Nitta, and Srimoyee~Sen
for useful discussions.
       Research of author GB was supported in part by NSF Grants PHY1305891 and PHY1714042. Author TH was supported by JSPS Grants-in-Aid for Scientific Research No.~15H03663 and 18H05236; GB and TH were partially supported by the RIKEN iTHES Project and iTHEMS Program. MGA is supported by the U.S. Department of Energy, Office of Science, Office of Nuclear Physics under Award Number \#DE-FG02-05ER41375.  KF was supported by Japan Society for the Promotion of Science (JSPS) KAKENHI Grant No.\ 15H03652, 15K13479, and 18H01211. MT was supported by JSPS Grant-in-Aid
 for Scientific Research, Grant No.\ 16K05357.
       The authors are grateful to the Aspen Center for Physics, supported by NSF Grant PHY1607611, where part of this research was done.

\bibliography{FluxTubes}{}
\bibliographystyle{apsrev4-1}

\end{document}